\documentclass{Interspeech2024}
\usepackage{ifthen}
\setboolean{interspeechfinal}{true}

\title{Explainable Attribute-Based Speaker Verification}

\name[affiliation={1}]{Xiaoliang}{Wu}
\name[affiliation={1}]{Chau}{Luu}
\name[affiliation={1}]{Peter}{Bell}
\name[affiliation={1}]{Ajitha}{Rajan}

\address{
  $^1$University of Edinburgh, United Kingdom}
\email{x.wu-53@sms.ed.ac.uk, chau.luu@ed.ac.uk, peter.bell@ed.ac.uk, arajan@exseed.ed.ac.uk}

\keywords{Speaker verification, Explanation}

\usepackage{url}
\usepackage{color}
\usepackage{algorithm}
\usepackage{algpseudocode}
\usepackage{verbatim}
\usepackage{colortbl}
\usepackage{multirow}
\usepackage{xcolor}     
\usepackage{amsmath,amssymb}

\begin{document}

\maketitle

\begin{abstract}
This paper proposes a fully explainable approach to speaker verification (SV), a task that fundamentally relies on individual speaker characteristics. The opaque use of speaker attributes in current SV systems raises concerns of trust. Addressing this, we propose an attribute-based explainable SV system that identifies speakers by comparing personal attributes such as gender, nationality, and age extracted automatically from voice recordings. We believe this approach better aligns with human reasoning, making it more understandable than traditional methods. Evaluated on the Voxceleb1 test set, the best performance of our system is comparable with the ground truth established when using all correct attributes, proving its efficacy. Whilst our approach sacrifices some performance compared to non-explainable methods, we believe that it moves us closer to the goal of transparent, interpretable AI and lays the groundwork for future enhancements through attribute expansion.
\end{abstract}

\section{Introduction}

The recent surge in Explainable Artificial Intelligence (XAI) techniques marks a significant evolution in AI technology. Advancements in XAI -- aimed at making deep learning models more understandable -- are particularly notable in the fields of image processing and natural language processing~\cite{LIME,DLIME,LIMEtree,deeplift,intergratedgradients,grad-cam}. Such developments, coupled with the emergence of regulatory frameworks such as the EU's AI Act\footnote{https://artificialintelligenceact.eu/the-act/}, highlight the necessity of making AI decision-making processes more transparent and amenable to human inspection. Further, there are signs of pivotal shift in approach: as \cite{Abeywickrama_2023} argues it is no longer adequate to simply attach explainability to AI systems as an afterthought. Rather, explainability must be integrated into the foundations of AI system design, ensuring that it is as fundamental as other system specifications. 

This paper develops an explainability-first design to the AI task of automatic speaker verification (SV).  SV is one example of a family of AI techniques applied to the field of spoken language processing, of which perhaps the best-known is automatic speech recognition (ASR).   Because speech is a signal that inherently contains personal information, there has been concern that AI systems may exhibit varying performance for different groups of speakers, or else use speaker information in an inappropriate way. For example, \cite{Tatman2017EffectsOT,Koenecke2020} showed ASR to have higher error rates for particular ethnic groups, on account of pronunciation and dialect differences. 

The challenge of explainability is particularly problematic for SV.  In contrast to ASR, where it would be reasonable to design systems that operate independently of personal characteristics its users' voices, SV has a fundamental reliance on speaker characteristics \cite{chau2020} and is likely to make use of group-level speaker properties such as accent, age and gender when making decisions, in much the same way that natural human decisions are made.  This poses a unique challenge when applying XAI techniques to this task: in this work, we attempt to resolve the conundrum by designing an XAI approach for SV that \textit{does} make use of group-level speaker attributes, but does so in a fully transparent manner.

The use of group-level attributes in SV not entirely new.  In 2020, Luu et al.\cite{chau2020} aimed to boost performance by jointly training deep speaker embeddings on tasks related to speaker attributes, such as age and gender, along with speaker classification. However, they did not integrate speaker attributes directly into the speaker verification process itself.  Unlike \cite{chau2020}, our aim is to expose these personal attributes and directly use them to generate SV decisions, creating a model that is as transparent as possible in its use of this characteristics.  We achieve this goal by using the idea of Concept-Bottleneck Model (CBM)\cite{CBM} that embeds an intermediate attribute bottleneck layer as the classifiers of the chosen attributes, using their predictions for final decision-making.   Our aim is not to produce the best possible SV accuracy scores, but rather, to demonstrate what can be achieved on this task in a transparent and  explainable setting. Source code available at { \small\textbf{\url{https://anonymous.4open.science/r/explainable-SV-E3C2}}}. 


We note that in the speech domain, post-hoc explanation techniques like LIME\cite{LIME} and SHAP\cite{SHAP} have perviously been applied to tasks such as ASR\cite{wu2023} and Phoneme Recognition\cite{wu2023trust}, identifying key audio segments. However, their visual, post-hoc nature lacks intuitiveness for SV task.

\section{Explainable Attribute-Based SV}
\label{sec: method}

\begin{figure*}[ht]
    \centering
    \includegraphics[width = \linewidth]{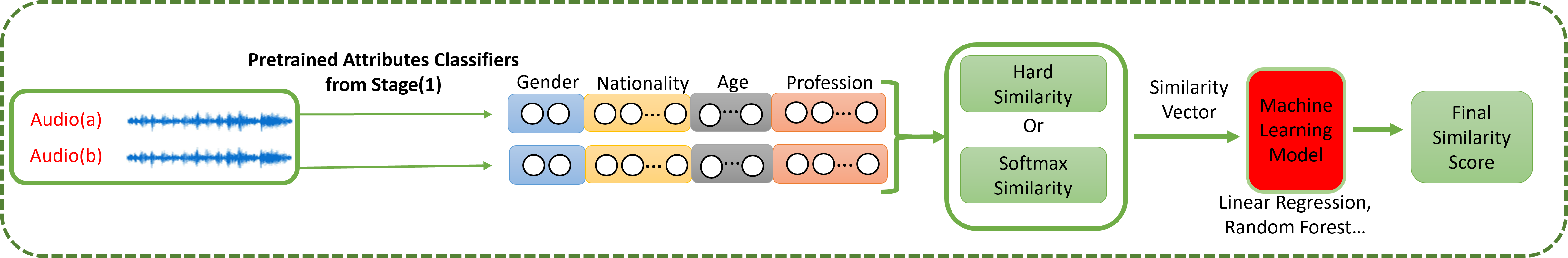}
    \caption{\texttt{Stage-2} of our SV system is shown. Pretrained classifiers from \texttt{stage-1} is used to extract attribute labels from pairs of audio samples. These attributes are then fed into a computation block that calculates a similarity vector for the corresponding pair of audio using hard or softmax similarity. The similarity vector is then used to train a \texttt{stage-2} Machine Learning Model, shown in red, which is the only component being trained during this stage. The output is the final similarity score, which quantifies the likelihood that the two audio inputs are from the same speaker.}
    \label{fig: explainableSV}
\end{figure*}



To develop a speaker verification (SV) system based on human-comprehensible attributes, we adapt the idea of the Concept-Bottleneck Model (CBM)\cite{CBM}, an approach that introduces an intermediate layer designed to explicitly learn and represent human-understandable concepts or attributes from input features.  In the next step, a predictor leverages these concept predictions to determine the final label. The utility of CBMs is well-demonstrated through their application in medical imaging, as highlighted by Koh et al.\cite{CBM}, where CBMs effectively identify critical health indicators, such as lung opacity and pleural effusion, from X-ray images prior to making a diagnosis.

Similarly, our explainable attribute-based SV also utilizes a two-stage training pipeline. The first stage involves constructing \texttt{stage-1} attribute classifiers for predicting attributes like gender, age, nationality, and profession from audio features, essential for distinguishing speakers. The second stage, as shown in the Figure~\ref{fig: explainableSV}, leverages the attributes generated by these classifiers to train a \texttt{stage-2} machine learning model to perform the speaker verification task.  We discuss the two stages in our pipeline below. 

\subsection{Stage-1: Attribute Classifiers}

In the first stage, we focus on training multiple classifiers, each dedicated to predicting a specific attribute. We use two distinct training approaches to optimize attribute prediction. 

\textbf{Xvector and ECAPA}: In the first approach, we use sophisticated speaker embeddings as inputs for training our \texttt{stage-1} attribute classifiers. These embeddings are derived from two pre-trained SV models: the Xvector architecture~\cite{xvector}, and the state-of-the-art ECAPA model~\cite{desplanques2020ecapa}. Using pretrained embeddings, our classifiers are able to analyze complex speaker characteristics in depth, leveraging the rich, informative embeddings produced. We emphasize that in the rest of this paper, both `ECAPA' and `Xvector' refer to \texttt{stage-1} attribute classifiers that take \emph{ECAPA embeddings} and \emph{Xvector embeddings} as inputs, respectively.

\textbf{Attributes Classification Technique (AC)}: In the second approach, we directly train \texttt{stage-1} attribute classifiers using Mel Frequency Cepstral Coefficients (MFCCs) as the primary input. The architecture details are in Section~\ref{sec: setup}. 
This approach is designed for direct analysis of audio signals, allowing our classifiers to efficiently process the raw audio data. 

These two approaches allow for flexibility within our classifiers, catering to different requirements of speaker identity analysis —- whether it needs analysis using sophisticated speaker embeddings or direct classification using raw audio features.

\subsection{Stage-2: Attribute-based SV}

After obtaining \texttt{stage-1} attribute classifiers, we start the second stage of training SV using attributes, shown in Figure~\ref{fig: explainableSV}, that comprises the steps described below. 

\textbf{Creation and Labeling of Audio Pairs:} We begin by generating a collection of audio pairs, categorizing them into positive pairs (from the same speaker) labeled as $1$, and negative pairs (from different speakers) labeled as $0$. This setup forms the groundwork for training our explainable models and is further discussed in Experiment Setup Section~\ref{sec: setup}. 

\textbf{Attribute Classifier Processing:}
Each audio in the pairs is analyzed using the \texttt{stage-1} attribute classifiers to extract attribute labels and their corresponding probability vectors from the last layer of the classifier. A similarity vector for each pair is then constructed, with dimensions equal to the number of selected attributes, to quantify the attribute-wise similarity. 

To compute the similarity vector, we first need to determine the similarity of the corresponding attribute labels between the audio pairs. 
To address this, we propose two methods.
\begin{itemize}
\item \textbf{Hard Label Similarity:} In this approach, the similarity is considered binary: 0 for different labels, 1 for same labels.  

\item \textbf{Softmax Label Similarity:}
Rather than a strict comparison with hard labels, this method compares probability vectors from the classifers' last layer. A probability vector reflects the likelihood of each potential class. 
For example, in nationality classification, a probability vector indicates the likelihood of the speaker belonging to each of several countries. We then calculate the cosine similarity between these vectors for each attribute. 

\end{itemize}

\textbf{Calculating Similarity and Constructing Vectors: }We then compute the similarity using either hard labels or softmax labels for each attribute, constructing a similarity vector $[s_{gen}, s_{nat}, s_{age}, s_{pro}]$ for each audio pair. Here, $s_{gen}$, $s_{nat}$, $s_{age}$, and $s_{pro}$ represent the similarity scores for gender, nationality, age, and profession, respectively.

\textbf{Training with Machine Learning Models:}
Each of the aforementioned similarity vectors serves as a training point that is fed into a machine learning model for training. We explore several machine learning models, namely Linear Regression, Random Forest and the Logistic Regression. 

In addition to these simple models, we also explore a simple neural network based model with two layers—a hidden layer and an output layer. The neural network takes the similarity vectors as input and processes them through the hidden layer, using a sigmoid activation function to capture complex relationships between attributes, finally outputting a similarity score for the audio pair.  

\section{Attributes and Datasets}

Given the accessibility of datasets and the availability of labels, we initially chose gender, age, and nationality as our main attributes for speaker verification, based on existing research\cite{chau2020} that highlights their influence on speech characteristics. The choice of attributes is also supported by insights from forensic phonetics~\cite{Jessen2007,Hansen2015}, where these attributes have been shown to be essential for identifying speakers in various contexts. 

Furthermore, we include profession, a previously unexplored attribute, in our set of considered attributes. Profession is defined as the career or occupation of individuals. 
Similar to gender and nationality, profession impacts speech patterns and is used as a proxy for sociolect of speaker characteristics. Sociolect is commonly used in forensic phonetics~\cite{Jessen2007,Hansen2015}, affecting speech through lexicon, syntax, and stylistic choices, shaped by an individual's educational and professional background.
To confirm our hypothesis that profession is representative of speaker characteristics, we verify in Section~\ref{sec: result} that profession is predictable from audio input. 
Finally, profession labels are not difficult to source. 
Since speakers in Voxceleb are well-known individuals, we are able to source this profession information from Wikipedia, acquiring 49 distinct profession labels that included politicians, actors, singers among others.

\textbf{Datasets: } We use VoxCeleb, augmented
in the standard Kaldi\cite{kaldi} fashion with babble, music and background noises along with reverberation, for speaker's gender, nationality, and profession, and the SCOTUS corpus, for age. Labels for gender, nationality, and age come from~\cite{chau2020}. 

\section{Experimental Setup}
\label{sec: setup}


\textbf{Stage-1}: For training \texttt{stage-1} attribute classifiers with Xvector and ECAPA embeddings, we employ two hidden layers leading to class projection, with Leaky ReLU activation functions. Alternatively, when training AC on MFCCs, our architecture consists of several Time-Delay Neural Network (TDNN) layers, a pooling layer, two fully connected layers, and a final softmax layer for label classification. 

Training parameters: for every \texttt{stage-1} attribute classifier, we run 100,000 iterations with a batch size of 256. Stochastic gradient descent is used for optimization, starting with a learning rate of 0.2 and a momentum setting of 0.5.

\noindent \textbf{Stage-2}: We create a training dataset from a random selection of 160 speakers from VoxCeleb2, generating 80,000 positive (same-speaker) and 80,000 negative (different-speaker) trials, totaling 160,000 trials. This dataset is utilized to train our \texttt{stage-2} machine learning models. 

\noindent \textbf{Test Set}: \texttt{Stage-1} attribute classifers and \texttt{stage-2} machine learning models are evaluated on the VoxCeleb 1 test set, which consists of 40 speakers and encompasses 37,720 trials.

\section{Metrics}

We use two metrics: Accuracy and Equal Error Rate (EER). \textbf{Accuracy} reflects the precision of \texttt{stage-1} attribute classifiers in identifying attributes correctly--the higher, the better. \textbf{EER} is a performance measure used in SV to determine the threshold value where the false acceptance rate (FAR) equals the false rejection rate (FRR). A lower EER indicates better performance.

\section{Results and Discussions}
\label{sec: result}
We present results for the following evaluations on Voxceleb1 test set: 1. We compare the three \texttt{stage-1} attribute classifiers and their impact on the downstream SV task; 2. We examine the impact of using Hard Labels versus Softmax Labels during \texttt{stage-2}; 3. We compare the relative importance of the four attributes for the SV task within our framework. 

\begin{table}[ht]
\centering
\begin{tabular}{|c|c|c|c|c|}
\hline
            & Random & Xvector & ECAPA & AC                       \\ \hline
Gender      & 0.50   & 0.99        & 0.99  & { 0.99} \\ \hline
Nationality & 0.32   & 0.72        & 0.70  & {\color[HTML]{FE0000} 0.75} \\ \hline
Profession  & 0.13   & 0.56        & 0.65  & {\color[HTML]{FE0000} 0.67} \\ \hline
Age         & 0.17   & 0.66        & 0.73  & {\color[HTML]{FE0000} 0.78} \\ \hline
\end{tabular}

\caption{Accuracy of three sets of \texttt{stage-1} attribute classifiers—Xvector, ECAPA, and AC—across four attributes (gender, nationality, profession, and age).}
\label{tab: acc}
\end{table}

\begin{table*}[ht]
\centering
\resizebox{140mm}{15.5mm}{
\begin{tabular}{|c|cccccccc|}
\hline
\multicolumn{1}{|l|}{} & \multicolumn{1}{c|}{}                     & \multicolumn{3}{c|}{\cellcolor[HTML]{FFCCC9}Softmax Labels}                                                                                                                                                                            & \multicolumn{3}{c|}{\cellcolor[HTML]{ECF4FF}Hard Labels}                                                                                                                                                                               &        \\ \hline
                       & \multicolumn{1}{c|}{\textbf{Groundtruth}} & \multicolumn{1}{c|}{\cellcolor[HTML]{FFCCC9}{\color[HTML]{333333} Xvector}} & \multicolumn{1}{c|}{\cellcolor[HTML]{FFCCC9}{\color[HTML]{333333} ECAPA}} & \multicolumn{1}{c|}{\cellcolor[HTML]{FFCCC9}{\color[HTML]{333333} AC}}   & \multicolumn{1}{c|}{\cellcolor[HTML]{ECF4FF}{\color[HTML]{333333} Xvector}} & \multicolumn{1}{c|}{\cellcolor[HTML]{ECF4FF}{\color[HTML]{333333} ECAPA}} & \multicolumn{1}{c|}{\cellcolor[HTML]{ECF4FF}{\color[HTML]{333333} AC}}   & Random \\ \hline
Linear Regression      & \multicolumn{1}{c|}{0.15}                 & \multicolumn{1}{c|}{\cellcolor[HTML]{FFCCC9}{\color[HTML]{333333} 0.22}}      & \multicolumn{1}{c|}{\cellcolor[HTML]{FFCCC9}{\color[HTML]{333333} 0.21}}    & \multicolumn{1}{c|}{\cellcolor[HTML]{FFCCC9}{\color[HTML]{FE0000} 0.20}} & \multicolumn{1}{c|}{\cellcolor[HTML]{ECF4FF}{\color[HTML]{333333} 0.36}}      & \multicolumn{1}{c|}{\cellcolor[HTML]{ECF4FF}{\color[HTML]{333333} 0.26}}    & \multicolumn{1}{c|}{\cellcolor[HTML]{ECF4FF}{\color[HTML]{FE0000} 0.25}} & 0.50   \\ \hline
Random Forest          & \multicolumn{1}{c|}{0.15}                 & \multicolumn{1}{c|}{\cellcolor[HTML]{FFCCC9}{\color[HTML]{333333} 0.22}}      & \multicolumn{1}{c|}{\cellcolor[HTML]{FFCCC9}{\color[HTML]{FE0000} 0.18}}    & \multicolumn{1}{c|}{\cellcolor[HTML]{FFCCC9}{\color[HTML]{333333} 0.21}} & \multicolumn{1}{c|}{\cellcolor[HTML]{ECF4FF}{\color[HTML]{333333} 0.27}}      & \multicolumn{1}{c|}{\cellcolor[HTML]{ECF4FF}{\color[HTML]{FE0000} 0.23}}    & \multicolumn{1}{c|}{\cellcolor[HTML]{ECF4FF}{\color[HTML]{333333} 0.26}} & 0.50   \\ \hline
Logistic Regression    & \multicolumn{1}{c|}{0.15}                 & \multicolumn{1}{c|}{\cellcolor[HTML]{FFCCC9}{\color[HTML]{333333} 0.21}}      & \multicolumn{1}{c|}{\cellcolor[HTML]{FFCCC9}{\color[HTML]{FE0000} 0.20}}    & \multicolumn{1}{c|}{\cellcolor[HTML]{FFCCC9}{\color[HTML]{FE0000} 0.20}} & \multicolumn{1}{c|}{\cellcolor[HTML]{ECF4FF}{\color[HTML]{333333} 0.29}}      & \multicolumn{1}{c|}{\cellcolor[HTML]{ECF4FF}{\color[HTML]{FE0000} 0.23}}    & \multicolumn{1}{c|}{\cellcolor[HTML]{ECF4FF}{\color[HTML]{FE0000} 0.23}} & 0.50   \\ \hline
Neural Network         & \multicolumn{1}{c|}{0.15}                 & \multicolumn{1}{c|}{\cellcolor[HTML]{FFCCC9}0.21}                             & \multicolumn{1}{c|}{\cellcolor[HTML]{FFCCC9}{\color[HTML]{FE0000} 0.18}}    & \multicolumn{1}{c|}{\cellcolor[HTML]{FFCCC9}0.20}                        & \multicolumn{1}{c|}{\cellcolor[HTML]{ECF4FF}0.36}                             & \multicolumn{1}{c|}{\cellcolor[HTML]{ECF4FF}{\color[HTML]{FE0000} 0.23}}    & \multicolumn{1}{c|}{\cellcolor[HTML]{ECF4FF}0.25}                        & 0.50   \\ \hline
Xvector-org            & \multicolumn{8}{c|}{0.035}                                                                                                                                                                                                                                                                                                                                                                                                                                                                                                           \\ \hline
ECAPA-org              & \multicolumn{8}{c|}{0.018}                                                                                                                                                                                                                                                                                                                                                                                                                                                                                                           \\ \hline
\end{tabular}}
\caption{EER of 4 \texttt{stage-2} machine learning models(Linear regression, Random Forest, Logistic Regression, Neural Network) using softmax labels and hard labels from three sets of \texttt{stage-1} attribute classifiers(Xvector, ECAPA, and AC).}
\label{tab: behaviours_explanation}
\end{table*}

\begin{table}[ht]
\centering
\resizebox{\columnwidth}{12.5mm}{
\begin{tabular}{|c|c|c|c|c|c|}
\hline
            & Random & Xvector & ECAPA & AC   & Groundtruth \\ \hline
Gender      & 0.50   & 0.40    & 0.34  & 0.34 & 0.36        \\ \hline
profession  & 0.50   & 0.29    & 0.26  & 0.26 & 0.25        \\ \hline
Nationality & 0.50   & 0.36    & 0.34  & 0.35 & 0.27        \\ \hline
Age         & 0.50   & 0.41    & 0.38  & 0.41 &             \\ \hline
All         & 0.50   & 0.20    & 0.18  & 0.21 & 0.15        \\ \hline
\end{tabular}}
\caption{EER when using gender-only, profession-only, nationality-only, age-only and all softmax labels from three sets of \texttt{stage-1} attribute classifiers. When all softmax labels are utilized, Random Forest is employed as the \texttt{stage-2} model.}
\label{tab: eer_seperate}
\vspace{-10pt}
\end{table}

\subsection{Comparison of \texttt{Stage-1} attribute classifiers:}
Table~\ref{tab: acc} compares the accuracies achieved with different \texttt{stage-1} attribute classifiers, namely Xvector, ECAPA, and AC for gender, nationality, profession, and age. The \texttt{Random} column is the baseline we compare against, that samples a given attribute based on its label distribution. 
We can see that all classifiers significantly outperform random guesses on all four attributes. Taking the profession attribute as an example, the random guess accuracy is 0.13, whereas our classifiers significantly outperform this with accuracies of 0.56 for Xvector, 0.65 for ECAPA, and 0.67 for AC, demonstrating a clear advantage over random guessing. This also goes to show that the profession attribute is predictable from voice recordings. 

Delving deeper into the comparison between Xvector, ECAPA, and AC, we notice that AC consistently outperforms the others across all attributes. This is because AC is trained directly with MFCCs, which retains more information than the pretrained embeddings used by Xvector and ECAPA classifers.
We also find that ECAPA generally surpasses Xvector, likely due to the use of more informative embeddings. 

Next, we explore how these classifiers support \texttt{stage-2} machine learning models. Table~\ref{tab: behaviours_explanation} shows the EER of 4 \texttt{stage-2} machine learning models--Linear regression, Random Forest, Logistic Regression and Neural Network--using softmax or hard labels from \texttt{stage-1} attribute classifiers: Xvector, ECAPA and AC.  
In Table~\ref{tab: behaviours_explanation}, the \texttt{Groundtruth} column shows the outcomes achieved using fully accurate labels for the attributes, representing the best possible outcomes with the given attribute set. The \texttt{Random} column, serving as our baseline, reflects outcomes from random guessing, offering a fundamental baseline of comparison. 

The red shaded area of Table~\ref{tab: behaviours_explanation} shows similarity vector computed using softmax labels and the blue shaded area is using hard labels. We find using the ECAPA classifier with softmax labels for similarity followed by Random Forest or Neural Network model achieves the best performance, with EER of 0.18. This outcome is slightly better than those obtained using Linear Regression and Logistic Regression, likely due to the ability of Random Forest and Neural Networks in capturing complex non-linear interactions among attributes. Also, it significantly surpasses the \texttt{Random} guess baseline of 0.50 and closely approaches the optimal \texttt{Groundtruth} result of 0.15, which shows the effectiveness of the ECAPA-Random Forest or ECAPA-Neural Network explainable attribute-based SV model.
It is worth noting that performance of AC is comparable to ECAPA across all four \texttt{stage-2} machine learning models, owing to the similar accuracies in attribute prediction observed in \texttt{stage-1}, as seen earlier in Table~\ref{tab: acc}. In contrast, Xvector classifiers slightly underperform ECAPA and AC across all \texttt{stage-2} machine learning models. This also aligns with our findings for \texttt{stage-1} attribute accuracy in Table~\ref{tab: acc} with Xvector having the least attribute prediction accuracy.

Given the comparable performance of ECAPA and AC, we further explore the scenarios in which these 2 \texttt{stage-1} attribute classifiers tend to make mistakes. In particular, when both are followed by a Random Forest (RF) \texttt{stage-2} machine learning model and tested on Voxceleb1 test set, with over 37,000 audio pairs, ECAPA incorrectly classified approximately 6,600 pairs, while AC misclassified 7,700 audio pairs. Among the misclassifications, we find ECAPA and AC both misclassify a shared 3,400 audio pairs. Focusing on the unique errors (3,200 for ECAPA-RF and 4,300 for AC-RF), we aim to identify error trends in attributes or speakers. 
Analysis shows no significant differences between ECAPA and AC in terms of gender and profession predictions. However, for nationality, ECAPA had a slightly higher error rate for speakers from India and Mexico. 
Additionally, we examined error probabilities of ECAPA and AC for each speaker in the test set. We find for speakers \texttt{id10301} and \texttt{id10292}, who belong to India and USA with profession all actors,  ECAPA's error probabilities are lower than AC. Audio samples for these speakers is accompanied by considerable background noise and goes to show that ECAPA embeddings are more robust to noise than AC, reducing impact downstream on the \texttt{stage-2} machine learning models.   
Results for the speaker errors with ECAPA and AC are shown in Section 1 of the supplementary material pdf.

It is worth noting that the optimal performance of our explainable SV model (0.18) and the best achievable performance (0.15) with the current set of four attributes is higher than traditional ECAPA and Xvector systems with EERs of $0.035$ and $0.018$, respectively.  This limitation comes from the limited number and scope of available attributes. We believe that expanding the attribute set in the future with more relevant features will address the performance shortfall, while maintaining its explainability advantages.


In summary, regardless of which \texttt{stage-2} machine learning models is employed, AC and ECAPA demonstrate comparable performance on the SV task. Notably, ECAPA, when followed by either a Random Forest or a Neural Network, exhibits a slight advantage. Furthermore, ECAPA is more adept for use in noisy environments, showcasing broader applicability. Given these findings, we recommend ECAPA as the more favorable option. 

\subsection{Softmax versus Hard Label Similarity in Stage-2:}
Across all \texttt{stage-1} attribute classifiers and \texttt{stage-2} machine learning models, we find similarity using softmax labels produces lower EER than hard labels. 
We find the nuanced understanding of category probabilities reflected in softmax labels is subsequently useful for the \texttt{stage-2} machine learning models for computing the final similarity score, as witnessed by the lower EER in Table~\ref{tab: behaviours_explanation}. The adoption of softmax labels also narrows the EER performance gaps between \texttt{stage-1} attribute classifiers for the different \texttt{stage-2} machine learning models, reducing the disparity in EER performance from a maximum of $0.11$ difference (Xvector-Linear Regression versus AC-Linear Regression) to $0.02$ with the extra information on category probabilities. 



\subsection{Importance of Attributes}
We assess and compare the EER achieved when using a single attribute -- gender-only, profession-only, nationality-only and age-only softmax labels from the three \texttt{stage-1} attribute classifiers against random guessing that samples a given attribute based on its label distribution and ground truth attribute performance. Results are presented in Table~\ref{tab: eer_seperate}. We also show the EERs when considering \texttt{All} attributes with random forest \texttt{stage-2} best performing model for the sake of comparison. 

Examining the groundtruth EERs for each of the attributes in Table~\ref{tab: eer_seperate}, we find profession (0.25) and nationality (0.27) as the best performing attributes, surpassing gender (0.36). This suggests that, with accurate predictions, both profession and nationality are useful indicators of speaker identity.

With softmax labels from \texttt{stage-1} attribute classifiers, profession consistently outperforms other attributes across all \texttt{stage-1} classifiers. This shows profession's value in the SV task. Afer profession, nationality and gender have comparable importance based on their EERs. We believe the higher EER of nationality compared to profession is due to the sparsity of nationalities in the Voxceleb1 test dataset. 
Gender, with just two labels, offers less discriminative power but remains relevant to SV. Conversely, age is found to be less impactful (Xvector: 0.41, ECAPA: 0.38, AC: 0.41). 
\texttt{Stage-2} machine learning models, such as Linear Regression and Random Forest, also provide importance scores for the four attributes. We find the trend in the significance of these attributes aligns with the observations in Table~\ref{tab: eer_seperate}.

Overall, our analysis suggests prioritizing profession and nationality attributes for speaker verification with our test dataset, as they consistently show lower EERs and higher effectiveness compared to gender and age.

\section{Conclusion}
Our research shows a significant step towards transparent and interpretable AI in speaker verification. We develop a two-stage attribute-based explainable framework, beginning with training \texttt{stage-1} attribute classifiers, followed by using \texttt{stage-2} machine learning models on attributes labels for the SV task. While illustrating the potential for explainable SV, our system shows a performance sacrifice
 due to the current limited attribute set. We will aim to expand these attributes in our future work, seeking to improve performance while providing explainability, promising a new direction for SV systems.

\bibliographystyle{IEEEtran}
\bibliography{mybib}

\end{document}